\documentclass[10pt]{article}
\usepackage{amssymb,amsmath}
\usepackage{graphicx}
\usepackage{color}
\usepackage{xcolor}
\usepackage{curves,multicol}
\usepackage{geometry}
\usepackage{times}
\usepackage{bm}
\usepackage{ulem}
\usepackage{hyperref}
\usepackage{ucs}
\usepackage[utf8x]{inputenc}
\usepackage[intoc]{nomencl}
\usepackage{float}
\usepackage{subfig}
\usepackage{units}
\usepackage{titlesec}
\usepackage{mathptmx}
\usepackage{lscape}
\usepackage{nopageno}
\usepackage[labelfont=bf]{caption}
\usepackage{lipsum}   
\usepackage[none]{hyphenat} 
\usepackage{breqn}

\makeatletter

\renewcommand{\section}{\@startsection
{section}
{1}
{0mm}
{-\baselineskip}
{0.5\baselineskip}
{\normalfont\bfseries\MakeUppercase}} 

{

\renewcommand{\subsection}{\@startsection
{subsection}
{2}
{0mm}
{0.5\baselineskip}
{0.25\baselineskip}
{\bfseries\normalsize}} 

\makeatother

\newcommand{\Fro}{\text{\textit{Fr}}}
\newcommand{\Rey}{\text{\textit{Re}}}

\usepackage{hyperref}

\begin{document}
\sloppy

\setcounter{secnumdepth}{-1} 

\vspace*{-1.0cm}
\begin{flushright} \vbox{
34$^{\mathrm{th}}$ Symposium on Naval Hydrodynamics\\
Washington DC, USA, June 26-July 1 2022}
\end{flushright}

\vskip0.65cm
\begin{center}
\textbf{\LARGE
The high-Re far wake of a slender body and the effects of density stratification \\[0.35cm]
}

\Large J. L. Ortiz-Tarin, S. Nidhan and S. Sarkar\\ 

University of California at San Diego, La Jolla, USA\\
\vspace*{0.25cm}

\end{center}

\begin{multicols*}{2}

\section{Abstract}
The high-Reynolds-number wake of a slender body with a turbulent boundary layer is investigated using a hybrid simulation. The wake generator is a 6:1 prolate spheroid and the Reynolds number based on the diameter $D$ is $Re = 10^5$. The transition of the unstratified wake to a state of complete self-similarity is investigated by looking for the first time into the far field of a slender body wake. Unlike bluff-body wakes, here the flow is not dominated by vortex shedding in the near wake. Instead, the recirculation region is very small, the near wake is quasi-parallel and is characterized by the presence of broadband turbulence. The simulations show that these particularities alter the decay rate of the wake significantly and challenge the notion of universality -- the idea that far from the body all turbulent wakes decay at the same rate. The influence of stratification in the wake is also studied at Froude numbers $\Fro=U_\infty/ND=2$ and $10$.  The results imply that the onset of buoyancy effects depends on the wake generator and that the far wake retains information of the initial conditions.  

\section{Introduction}
\label{SECintroduction}

Due to their low drag coefficients, slender bodies are extensively used in aerospace and naval applications. Multiple studies have described the flow around these bodies focusing on the drag force, the boundary layer and the flow separation \cite{Wang1970, Costis1989, Wang1990, Chesnakas1994, Fu1994, Constantinescu2002, Wikstrom2004}. However, despite their presence in many underwater applications, only a few studies have looked into the wake of a slender body \cite{Chevray1968, Jimenez2010, Posa2016, Kumar2018}. The near wake of a slender body with a turbulent boundary layer is characterized by having a small recirculation region, absence of strong vortex shedding \cite{Jimenez2010, Kumar2018}, and is thin relative to
 the wake of bluff bodies. These particular features lead to interesting effects further downstream: despite having a smaller drag coefficient than bluff bodies, the defect velocity ($U_d=U_\infty - U$) of the slender body wake can be larger than that of a bluff body, also the turbulent kinetic energy of the wake shows an off-center radial peak close to the location where the turbulent boundary layer separates on the body.

Studying slender body wakes at high $\Rey$ is challenging both in experiments and in simulations and all  previous work \cite{Chevray1968, Jimenez2010, Posa2016, Kumar2018} have only looked into the near wake. The maximum downstream distance studied in the wake of a slender body with a turbulent boundary layer is 20$D$. Additionally, the few studies that look into slender body wakes assume that the body moves in a neutral environment, where the density of the surrounding fluid is constant. However, in a realistic  marine environment, the effect of density stratification due to salinity and temperature  becomes relevant. 

\begin{table*}
\begin{center}
\begin{tabular}{lccccccccccc}
    Case  & Type &$Re$   &   $Fr$ &  $L_r$ & $L_{\theta}$ & $L^-_x$ & $L^+_x$ & $x_{extr.}$& $N_r$ & $N_{\theta}$ & $N_{x}$ \\[3pt]
       1   & body-inclusive &$10^5$ & ~~$\infty$~ & 5 & $2\pi$ & 8 & 15 & 6 & 746 &512 & 2560\\
       2   & body-inclusive &$10^5$ & ~~10 & 60 & $2\pi$ & 20 & 30 & 9 & 848 & 512 & 3072\\
       3  & body-inclusive & $10^5$ & ~~$2$ & 60 & $2\pi$ & 20 & 30 & 9 & 848 &512 & 3072\\
       4  & body-exclusive &$10^5$ & ~~$\infty$~ & 10 & $2\pi$ & - & 80 & 6& 479 &256 & 4608\\
       5  & body-exclusive &$10^5$ & ~~10 & 57 & $2\pi$ & - & 89  & 9& 619 & 256 & 4608\\
       6  & body-exclusive &$10^5$ & ~~$2$ & 57 & $2\pi$ & - & 89 & 9& 619 & 256 & 4608\\

  \end{tabular}
\end{center}
 \caption{Simulation parameters of the body-inclusive and body-exclusive simulations. $L_x^-$ and $L_x^+$ are the upstream and downstream distances. $x_{extr.}$ is the extraction location of the BI simulation that is fed into the BE simulation.} 
 \label{tab:bi}
\end{table*}

In the present work we simulate the wake of a slender body with a turbulent boundary layer and at zero angle of incidence using a hybrid method. We summarize results from  our simulations of the unstratified wake \cite{Ortiz2021} and also discuss ongoing work on density-stratified wakes. The Reynolds number is set to $Re=U_\infty D/\nu=10^5$. This is the first study that looks into the slender body far wake with a domain that spans $80D$ downstream. In this domain, we can study the transition of the unstratified flow to full self-similarity and test the validity of the classic wake decay exponents \cite{Tennekes1972,George1989,Pope2000}. 

Apart from the unstratified wake, we also simulate two levels of stratification: $\Fro=2$ and $10$.  Here, $\Fro = U_\infty/ND$ is defined using the relative velocity ($U$) between the body and the ambient, $D$ the minor-axis diameter of the spheroid,  the buoyancy frequency ($N$) where $N^2 = -(g/\rho_0) d\rho_b/dz$ is determined by the background density $\rho_b (z)$. Density stratification suppresses vertical motions, triggers the formation and sustenance of coherent structures, and leads to the radiation of internal gravity waves. More importantly, in a stratified environment the wake of a submersible lives longer than in a neutral environment, i.e., it takes more time for the flow disturbance to die out \cite{Spedding2014}. Whereas most of the previous work in the stratified wake literature has been devoted to the wake of bluff bodies, these simulations allow us to quantify the effects of ambient stratification in the wake of a slender body at a high Reynolds number. We are particularly interested in the following questions: (i) how do the features of a slender body in the intermediate and far wake compared with that of a bluff body? (ii) how does the wake turbulence interact with the boundary layer turbulence under the effects of buoyancy? (iii) how do these findings connect with the previous literature that looks in the far wake of bluff bodies?

\section{Methodology}
\label{SECmethods}

The study of turbulent wakes has always been limited by the size of the computational and experimental domain. Wakes reach self-similar states far from the wake generator and, as the downstream distance increases, measuring weaker turbulent quantities becomes more and more challenging. Additionally, in the case of stratified wakes, the location behind the body at which buoyancy starts affecting the flow is proportional to $\Fro$. This implies that the difference between weakly-stratified and unstratified wakes is only revealed very far downstream. 

To reduce the cost of studying wakes in very long domains, temporal simulations have been used widely \cite{Gourlay2001,Dommermuth2002, Brucker2010}. In temporal simulations, the wake generator is not included in the domain and a reference frame that moves with the mean flow velocity is used. The streamwise variation of the mean flow is assumed to be negligible and periodic boundary conditions are imposed so that the cost of the simulation can be reduced. The flow initialization is done by superposing a fluctuation velocity profile over a self-similar mean velocity profile. Although illuminating and affordable, the temporal simulations with artificial initial conditions can not capture body generated lee waves, near-wake buoyancy effects or the vortex shedding from the body. 

To circumvent these limitations, body inclusive (BI) simulations, which are computationally costly, have been used \cite{Ortiz-tarin2019,nidhan2019dynamic, Chongsiripinyo2020,Nidhan2020,nidhan2022analysis}. Body inclusive simulations resolve the flow physics without restrictive underlying assumptions. However, due the high computational cost of resolving the wall region near the body, the domain size is limited and only the near and intermediate parts of the wake can be investigated.

Recently, VanDine et al. \cite{VanDine2018} presented a hybrid spatially-evolving model which generalizes the hybrid temporally-evolving model of Pasquetti \cite{Pasquetti2011} so as to circumvent most of the aforementioned problems. This type of simulation uses inflow conditions generated from a body-inclusive simulation and performs a separate spatially-evolving simulation without including the body, namely the body exclusive (BE) simulation. This procedure allows us to relax the natural stiffness of the far-wake problem by using different spatial and temporal resolutions appropriate for the BI and BE stages. The grid cell in the BE simulation, which has to be  sufficiently small to adequately resolve wake turbulence, is still much larger than the one required to resolve the boundary layer. The consequently large time step in the BE simulation leads to considerable savings in computational time without compromising on accuracy. The extraction location is determined based on the near-body physics. VanDine et al. \cite{VanDine2018} found that it is important to choose the extraction location downstream of the recirculation zone and where the defect velocity is small relative to the freestream. In the current simulations, the extraction locations for unstratified and stratified cases are $x/D = 6$ and $9$ respectively, significantly away from the recirculation bubble \cite{Ortiz2021}. Relative to the unstratified case, the BI domain is longer in the stratified cases and the simulation has a finer grid at the $x/D=9$ extraction location. Since the BE stratified simulations have larger radial domain size,  we decided to choose  the extraction location as $x/D=9$ instead of $x/D=6$ to reduce as much as possible the size of the BE domain. After the planes have been extracted, they are interpolated to a new grid and are fed at the proper rate (corresponding to advection by the freestream)  into the inflow boundary of the new simulation. 

Here we use a hybrid simulation to simulate the far wake of a 6:1 prolate spheroid at zero angle of incidence with a tripped boundary layer at a $Re=10^5$.


\begin{figure}[H]
\centering
 \includegraphics[width=7cm]{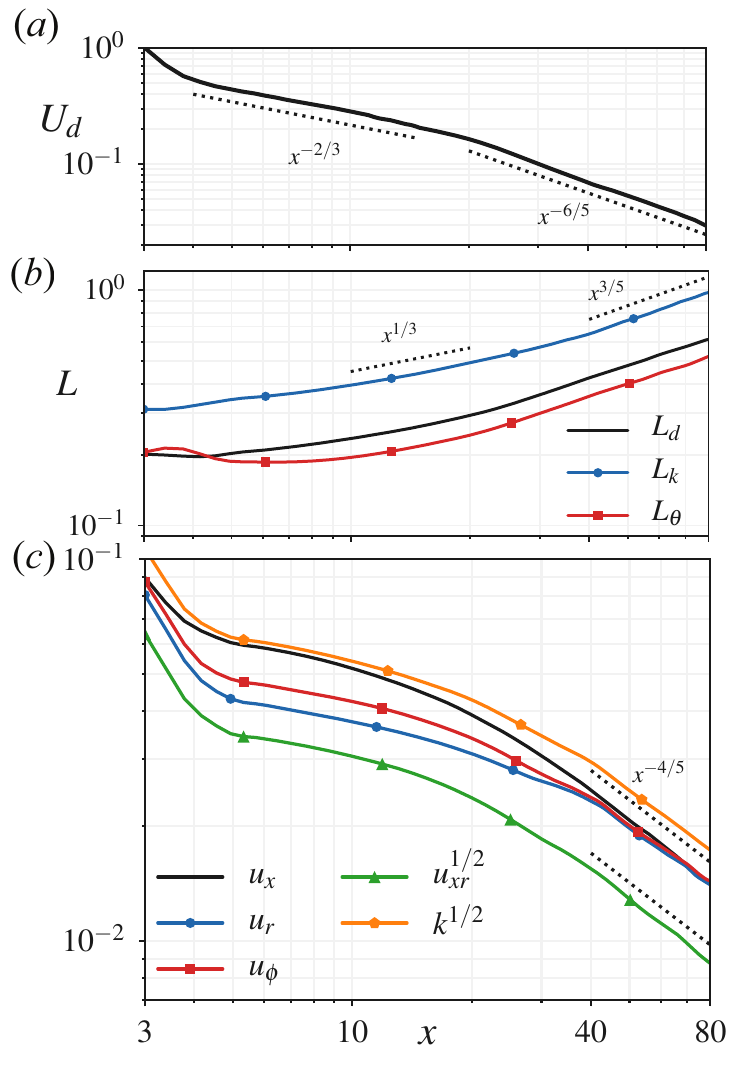}
 \caption{Streamwise evolution of wake statistics: (a) centerline defect velocity, (b) wake half-widths and (c) peak RMS velocities, Reynolds shear stress and TKE.}
\label{fig:1}
\end{figure}

Both the BI and BE simulations employ a Navier-Stokes finite differences solver on a cylindrical staggered grid. The solver advances in time using a fractional step method with a time advance that combines a third-order low storage Runge-Kutta method with the Crank-Nicolson method. The azimuthal periodicity of the cylindrical domain can be leveraged to reduce the cost of solving the pressure Poisson problem. To simulate the subgrid stress, the wall-adapting local eddy-viscosity (WALE) model of \cite{Nicoud1999} is employed. The WALE model enables us to properly capture the near-wall dynamics of the turbulent boundary layer (TBL) over the body in an economical fashion.
 In the wake, $\nu_t/\nu<1$, where $\nu_t$ and $\nu$ are subgrid and fluid kinematic viscosity, confirming excellent resolution of the LES for both BE and BI simulations. The body is represented using an immersed boundary method by \cite{Balaras2004}. The boundary layer is tripped by introducing a numerical disturbance at the surface of the body. Specifically, tripping is achieved by introducing a bump of radius $0.002D$ (15 wall units) at 0.5D distance from the nose of the spheroid. To avoid spurious reflection of the internal gravity waves, a sponge layer is added to the inlet, the outlet and the lateral wall of the domain. Additional details of the solver can be found in \cite{Ortiz-tarin2019, Chongsiripinyo2020}. The details of the resolution and domain size are given in table \ref{tab:bi}. For simplicity, two coordinate systems are employed. When the coordinate system is cylindrical, the radial, tangential and streamwise components are denoted by $\{r,\theta,x\}$. Cartesian coordinates are denoted by $\{x,y,z\}$, where $x$ is the streamwise coordinate, $y$ is the spanwise coordinate and $z$ is the vertical coordinate. The velocities are normalized by the freestream velocity $U_\infty$ and the lengths by the body diameter $D$.  

\section{Results}

\subsection*{The unstratified wake decay}

\begin{figure}[H]
	\centering
	\includegraphics[width=8cm]{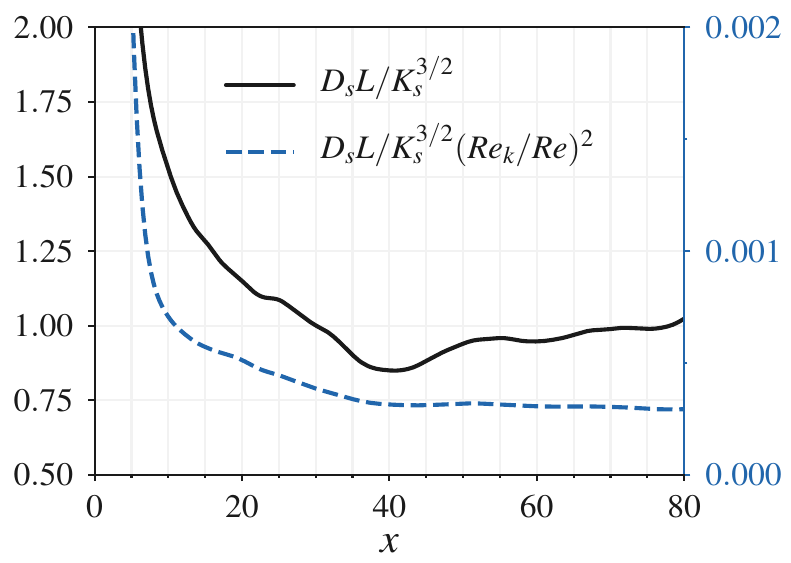}
	\caption{Peak turbulent dissipation ($D_s$) normalized with classical and with non-equilibrium estimates. $K_s$ corresponds to the maximum value of TKE ($k$).}
	\label{fig:2}
\end{figure}

In the unstratified wake, the decay of the defect velocity $U_d(x)$ exhibits two regions. As shown by \ref{fig:1}(a), the region  before $x \approx 20$,  where $U_d(x)\sim x^{-2/3}$, is followed by a second region where $U_d(x)\sim x^{-6/5}$. The results before $x \approx 20$ match previous simulations and experiments \cite{Chevray1968, Jimenez2010, Posa2016, Kumar2018}.  The following second decay rate has not been reported before since all previous studies have  reached up to  only $x\approx 20$ in the downstream direction. Different measures of the wake width are shown in figure \ref{fig:1}(b). $L_d$ is defined such that $U(L_d)=\frac{1}{2}U_d$, and  $L_k$ is defined analogously  using the TKE as $k(L_k)=\frac{1}{2}k(r=0)$. The displacement thickness ($L_\theta$)  is defined by
$L_\theta^2={U_d}^{-1}\int_0^\infty (U_\infty-U)rdr$. All three measures of wake width show similar wake power laws and they are consistent with conservation of momentum. Figure \ref{fig:1}(c) shows the decay of turbulent quantities as the flow evolves. In the region where $U_d \sim x^{-6/5}$, the turbulent quantities show a decay of $\sim x^{-4/5}$.

In the near wake region ($x \lessapprox 20$), the classic self-similar turbulence scalings ($k \sim  u_{xr} \sim U_d^{2}$ and  $\varepsilon \sim k^{3/2}/L$) that lead to the asymptotic $U_d\sim x^{-2/3}$ decay do not hold. Here, $u_{xr}$ corresponds to the peak value of the Reynolds shear stress term. The variables have not reached full self-similarity \cite{Ortiz2021} and the observed $U_d\sim x^{-2/3}$ is likely a short-lived coincidence with the classic wake result. 

After $x=40$, profiles of $k$ and $\varepsilon$ exhibit self-similarity \cite{Ortiz2021} (not shown here for brevity). The transition to the observed far-wake decay law of $U_d\sim x^{-6/5}$ is consistent with the observed non-equilibrium scaling of dissipation $\varepsilon~\sim(\Rey/\Rey_k)^2k^{3/2}/L$, see figure \ref{fig:2}. In the non-equilibrium scaling of dissipation, $Re$ is the global Reynolds number and $\Rey_k=\sqrt{k}L/\nu$ is the local turbulence Reynolds number. This non-equilibrium scaling has been previously observed in bluff-body wakes, jets and unsteady decaying turbulence \cite{Nedic2013, Dairay2015,Chongsiripinyo2020, Cafiero2019, Goto2016}. In the latter, it has been shown to be connected to the imbalance between the turbulent energy entering the flow through the large-scales and the dissipation occurring at the small-scales. 

\begin{figure}[H]
	\centering
	\includegraphics[width=8.0cm]{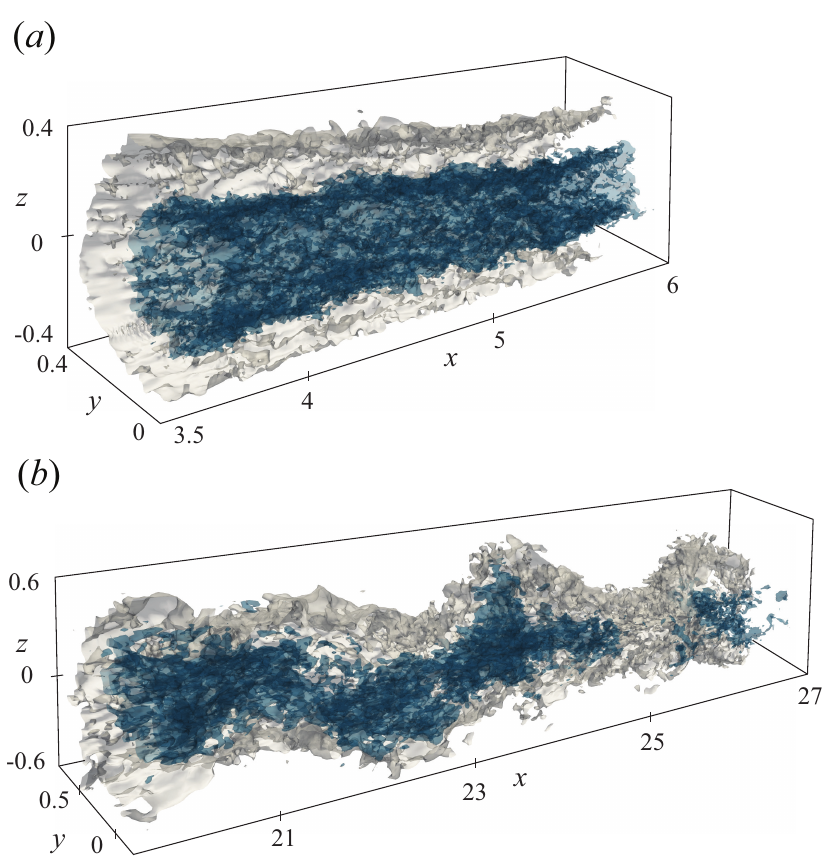}
	\caption{Isocontours of instantaneous streamwise velocity: (a) the near wake, white 0.97$U_\infty$, blue 0.7$U_\infty$, and (b) the intermediate wake, white 0.97$U_\infty$, blue 0.85$U_\infty$.}
	\label{fig:3}
\end{figure}

In previous bluff-body studies, the exponent in the non-equilibrium scaling of dissipation was $n=1$, while here $n = 2$. This difference is possibly related to a difference in the mechanism that sustains  large-scale wake motions. Here, the imprint of vortex shedding from the body is weak, large-scale coherent structures are only observed late in the wake, and the TKE and dissipation profiles show an off-center peak whereas, in bluff-body wakes such as those studied by Dairay et al. \cite{Dairay2015} and Nidhan et al. \cite{Nidhan2020}, the maxima occur at the centerline. Further investigation is still required for a more comprehensive understanding of the details of non- equilibrium scaling. Additionally, assessment of whether the turbulent dissipation transitions to the classic inertial estimate at larger downstream distances will require longer spatial domains.

 
After quantifying the decay rate in the near and far-wake regions, we investigate why the decay rate changes from the near to the far wake. The change in wake decay cannot be attributed to the low-$\Rey$ asymptotic result $U_d\sim x^{-1}$, since the local wake Reynolds numbers -- $\Rey=U_dL/\nu$, $\Rey_k=\sqrt{k}L/\nu$, and $\Rey_\lambda=\sqrt{k}\lambda/\nu$, where $\lambda$ is the Taylor microscale -- remain very large throughout the domain \cite{Ortiz2021}. It cannot be also attributed to the turbulence reaching equilibrium with the mean flow since $k$ does not scale with  $U_d^2$ (see figure \ref{fig:1}). We find that the decay rate transitions in the region where there is a structural change in the wake, e.g.,  figure \ref{fig:3}  shows that the  imprint of a sinuous single helix ($|m|=1$) azimuthal mode emerges only in the far wake (figure \ref{fig:3}(b)). In a bluff-body wake, the strong asymmetric shedding of the boundary layer leads to a dominant helical mode right after the recirculation bubble \cite{Nidhan2020, Berger1990, Johansson2006}. Here, the separated region behind the slender body is very small (about  $0.1D$), the near wake is quasi-parallel, and is dominated by the boundary layer turbulence leaving the body. The large-scale coherent helical structure is only visible further downstream preceding the change of decay rate. This transition, which is not observed in the wake of bluff bodies such as disks or plates, allows us to elucidate the possible role of the $|m|=1$ instability in the decay laws. 

The appearance of a large-scale coherent structure in the wake and the dominance of the $|m|=1$ mode precede the non-equilibrium scaling region. This finding supports the idea that there is a connection between the presence of coherent structures and the non-equilibrium scaling of dissipation. The $|m|=1$ contribution has been quantified through azimuthal power spectra (not shown here for brevity) and is found to become less dominant  towards the end of the computational domain ($x = 80$), indicating that the wake might eventually exhibit a further transition in its power laws, possibly to the classical power law. 

\subsection*{The effects of density stratification}

When a body moves in a stratified environment, there are two sources of internal gravity waves: the body and the fluctuating velocity in the wake. The body generates a steady (in the frame moving with the body)  coherent pattern of lee waves, whereas the fluctuating motions of the wake emit unsteady waves. The strength of the lee waves rapidly decays with increasing $Fr$ and  their influence is significant at $Fr\sim O(1)$. In this paper, only the body-generated lee waves are described, the analysis of the wake-generated waves is deferred to future work. The vertical velocity field induced by the lee wave pattern has two important effects on the flow. First, it strongly influences the separation of the boundary layer, directly affecting the turbulence content in the wake. Second, it modulates the decay of most of the wake properties.

\begin{figure}[H]
	\centering
	\includegraphics[width=8cm]{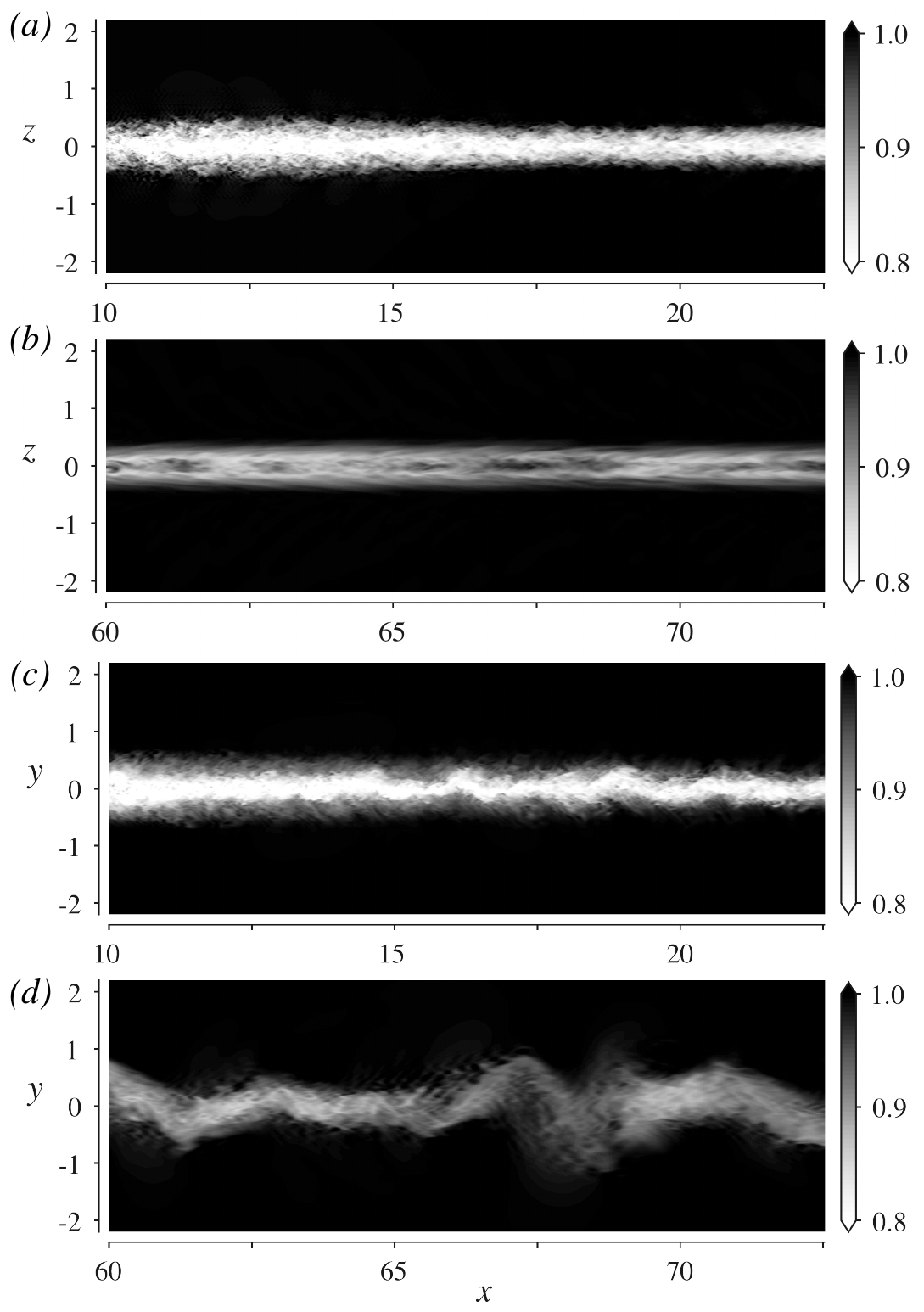}
	\caption{Isocountours of instantaneous streamwise velocity for the $\Fro = 2$ wake. Vertical planes in the near (a) and far wakes (b). Horizontal planes in the near (c) and far wakes (d).}
	\label{fig:4}
\end{figure}

\begin{figure}[H]
	\centering
	\includegraphics[width=8cm]{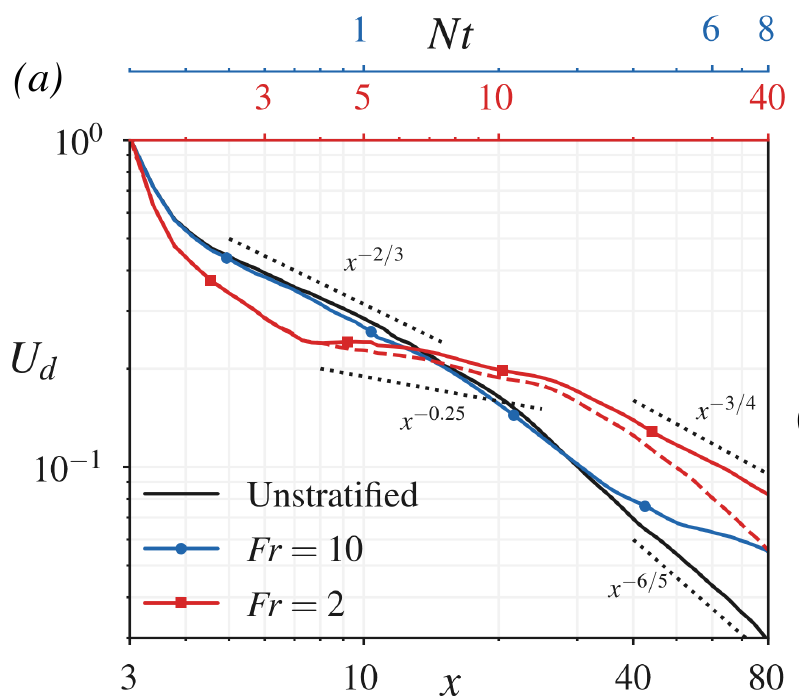}
	\caption{(a) Decay of the peak defect velocity as a function of the streamwise distance. The dashed line follows the defect velocity at the centerline, which at $Fr=2$ does not coincide with the peak.}
	\label{fig:5}
\end{figure}

Figure \ref{fig:4} reveals the main qualitative features of the wake at $Fr=2$. Buoyancy suppresses vertical motions and intensifies the horizontal motions. Consequently, stratified wakes tend to be typically wider than taller as can be seen by comparing figure \ref{fig:4}(b) to \ref{fig:4}(d). This interaction between the lee waves and the wake is particularly strong when the Froude number is close to a critical Froude number $Fr_c=AR/\pi$, where $AR$ is the body aspect ratio. When $Fr\approx Fr_c$, half the wavelength of the lee wave ($\lambda/D=2\pi Fr$) coincides with the length of the body and the size of the separation region is reduced. The flow is then in what is called the resonant or saturated lee wave regime, \cite{Hanazaki1988,Chomaz1992}. At low Reynolds numbers, this effect can lead to the relaminarization of the turbulent wake \cite{Ortiz-tarin2019}. For the 6:1 spheroid at hand, the $\Fro=2$ case is close to the critical $\Fro_c = AR/\pi \approx 1.9$ In the present case, figure \ref{fig:4}(a) reveals that, even at $Re=10^5$, the wake height is strongly modulated by the waves, although the wake is not relaminarized due to the high $\Rey$ of the flow. This oscillatory modulation leads to an interesting configuration in the intermediate wake ($x=20-40$) where the wake width $L_H$ is smaller than the wake height $L_V$, contrary to what is expected for a stratified wake. As the wake evolves, this configuration changes to the classic $L_H \gg L_V$. In the far-wake domain, the small-scale turbulence of the boundary layer has been dissipated and a triple-layer structure is observed in the vertical plane (figure \ref{fig:4}b) whereas large-scale horizontal motions are sustained in the horizontal (figure \ref{fig:4}d).    
\begin{figure}[H]
	\centering
	\includegraphics[width=8cm]{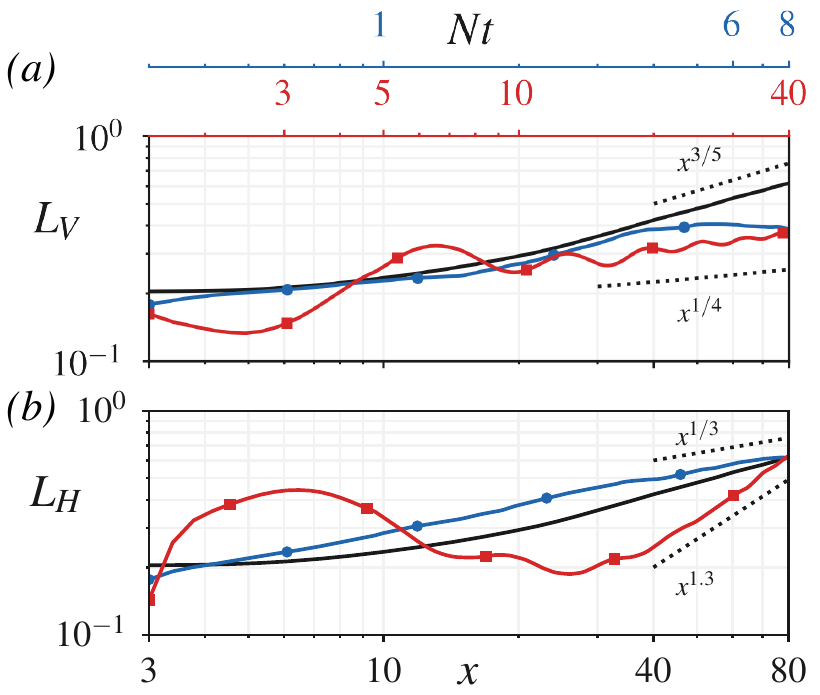}
	\caption{Evolution of (a) wake height $L_V$ and (b) wake width $L_H$ as a function of $x$ for $\Fro = \infty, 10$ and $2$ wakes. Color schemes same as in figure \ref{fig:5}.}
	\label{fig:6}
\end{figure}

\begin{figure}[H]
	\centering
	\includegraphics[width=8cm]{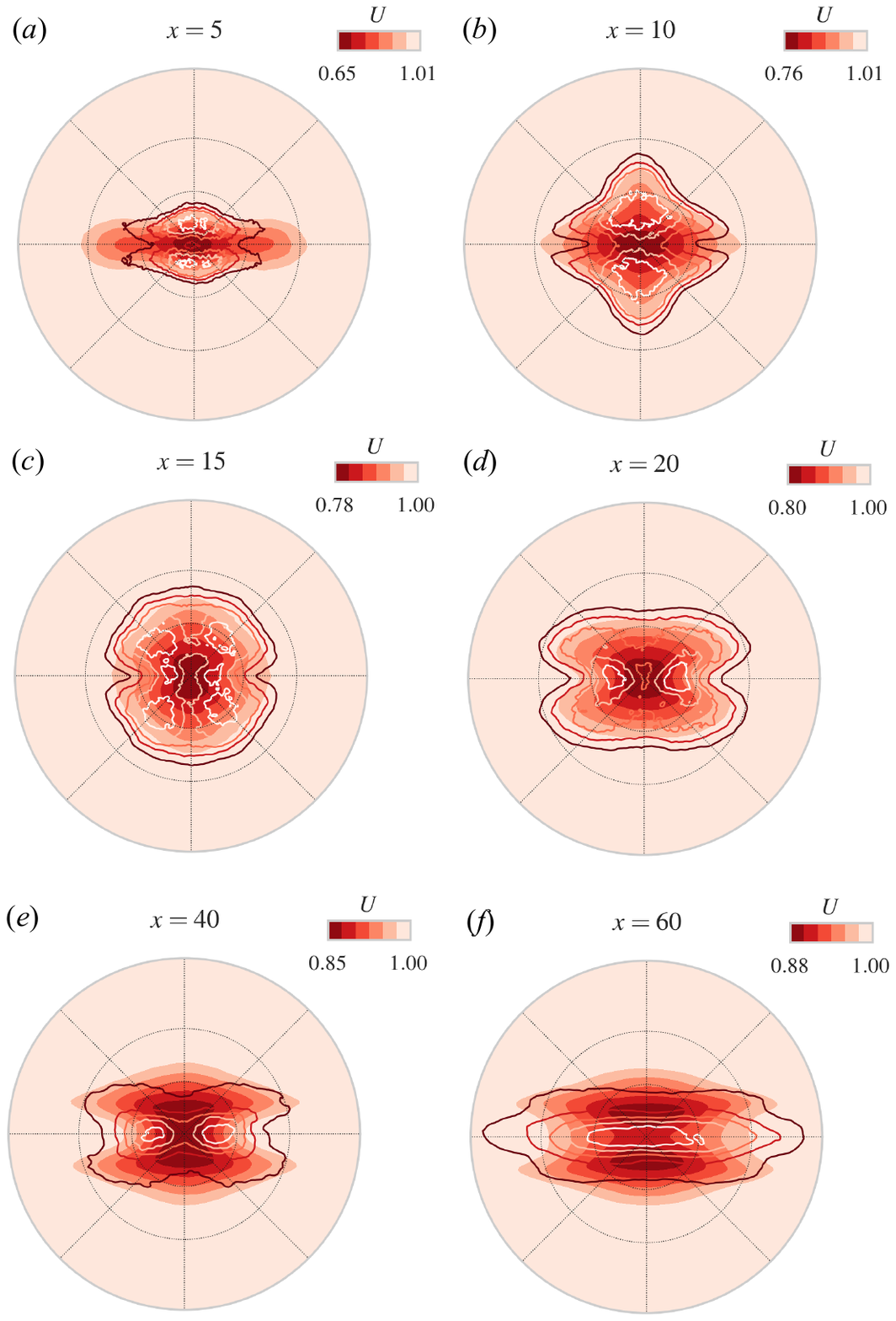}
	\caption{$\Fro=2$ wake. Contours of the mean streamwise velocity and the turbulent kinetic energy at different streamwise locations. Filled contours represent $U$. Unfilled isolines represent TKE contours plotted in increments of $15\%$ of the local TKE maximum.}
	\label{fig:7}
\end{figure}

The solid lines of figure \ref{fig:5} show the decay of the peak defect velocity. The $Fr=2$ wake shows a region of slow decay between $Nt=4-15$. The decay of $U_d\sim x^{-0.25} $ in this region coincides with the decay found in the experiments of Spedding \cite{Spedding1997}. This transitional decay is referred as the non-equilibrium regime (NEQ) where buoyancy effects becomes progressively stronger. The NEQ regime is followed by the Q2D regime with a faster power-law decay and quasi-two dimensional vortices. Note that this NEQ regime is not related to the non-equilibrium scaling of turbulent dissipation shown previously. In their experiments, Spedding \cite{Spedding1997} reports that the NEQ regime spans until $Nt\approx 40$.



Some numerical wake studies \cite{Brucker2010,Diamessis2011} have found that the span of the NEQ regime (which precedes the quasi 2D regime that exhibits a $x^{-3/4}$ wake defect power law) depends on the Reynolds number. 
 Here, in the  high-$\Rey$ slender body wake, we observe a transition to a $x^{-3/4}$ decay rate very early, at  $Nt=15$ instead of $Nt=50$. In ongoing work, we are investigating changes in the wake structure and turbulence associated with the transition in the power law. The red dashed line that separates from the solid red line in figure \ref{fig:5} follows the decay of the centerline value of the defect velocity. In the $Fr=10$ and $Fr=\infty$ wake, center and peak values coincide. In the $Fr=2$ wake, however, the topology of the wake is such that the peak defect velocity does not occur at the centerline beyond $x \approx 30$. As the $\Fro=2$ wake enters the Q2D regime, the mean wake becomes multilayered and the peak deficit shifts away from the centerline as also shown in figure \ref{fig:7}. This finding is consistent with previous numerical and experimental works on stratified wakes \cite{Brucker2010,Spedding1997}. Notice that, in the weakly stratified $Fr=10$ case, the defect velocity deviates from the unstratified case at $Nt=4$.


The wake width $L_H$ and height $L_V$ are calculated using  the locations  where the streamwise mean velocity deficit is reduced to half of its centerline value. Note that using the maximum deficit instead of the centerline deficit for the $\Fro=2$ wake does not change the qualitative trend in $L_V$. In quantitative terms, it leads to a  deviation of $\approx 5\%$, at maximum. In figure \ref{fig:6}(a), the height of the $Fr=2$ wake shows a very slow growth modulated with a wavelength of $\lambda/D=2\pi Fr$. This is the characteristic wavelength of the lee waves which originate at the body. In previous experiments with temporal wakes and blunt bodies, $L_V$ becomes locked once the NEQ regime starts, similar to what the $Fr=10$ wake shows at $Nt=4$. In this case, we observe a slow constant growth close to $L_V\sim x^{0.25}$. The evolution of $L_H$ shows a region of strong modulation by the gravity waves in  $3 < x < 30$ where the trend in $L_H$ opposes that in the evolution of $L_V$.  However, at $Nt\approx 15$  the horizontal growth rate increases significantly to a value of $x^{1.3}$. 

To further characterize the $Fr=2$ wake, figure \ref{fig:7} shows contours of mean velocity and turbulence at different streamwise locations. The $\Fro=10$ wake is not shown since its  topology is closer to the unstratified axisymmetric case. The mean can be well approximated by a vertically-squeezed two-dimensional Gaussian while the turbulence evolves from a bimodal distribution in the radial direction to a Gaussian.

The initial shape of the $Fr=2$ wake (figure \ref{fig:7}(a)) already shows the lack of axisymmetry. The shape of the mean velocity contour is flat and wide. TKE contours at $x = 5$ shows two off-center peaks, instead of an ellipsoidal distribution with a centerline peak. At $x=10$, the shape of the wake changes abruptly, driven by the lee waves. This is confirmed by the trends of $L_V$ and $L_H$, both of which follow the oscillations of the steady lee waves (figure 6). Between $x=10$ and $x=20$, the wake transitions to a `butterfly' shape reminiscent of the separation and wake patterns observed in \cite{Ortiz-tarin2019}. Note that, in this stage, the wake is thinner than taller $L_H<L_V$. Here, the TKE peaks are horizontally off-center. At this location, the horizontal sinusoidal motion shown in figure \ref{fig:4}(d) starts developing and, by $x=60$,  the turbulence no longer shows two distinct off-center peaks. Instead, a horizontal layer of turbulence is sustained between two vertically off-center peaks of mean velocity.  

Previous temporal simulations \cite{Gourlay2001,Redford2015} have shown that the mean and the turbulence can evolve differently. Indeed the effect of buoyancy is `felt' at different $Nt$ by the large and the small scales in the flow. The general trend is that, in the late wake, the turbulence occupies a smaller and smaller vertical fraction of the mean defect as time passes. In contrast, what we observe here is the combined effect of having a wake in saturated lee wave state and an off-center peak of TKE established by the turbulent boundary layer. These characteristics of the flow are not captured in temporal simulations (with artificial initial conditions) since they depend on the wake generator and the lee waves. The difference in the evolution of the mean and the turbulence is very substantial. 

\section{Conclusions}

In this study, we analyze the turbulent wake of a 6:1 prolate spheroid at $\Rey = 10^{5}$ and three body-based Froude numbers $\Fro = \infty, 2$, and $10$. Using the body-exclusive simulations, we access downstream distance up to $x/D = 80$ for all three cases, enabling the study of the far wake of a high-$\Rey$ slender body for the first time in the slender wake literature. For the unstratified case, unlike bluff-body wakes, the near wake is not dominated by the vortex shedding. Instead, the recirculation region is small, the near-wake is quasi-parallel, and the turbulence is broadband. These features have strong implications on how the wake statistics evolve. We find that the mean defect velocity $U_d$ decays as $U_d \sim x^{-6/5}$ for $30 < x/D < 80$, instead of the classical $U_d \sim x^{-2/3}$ scaling. Further analyses show that this scaling emerges due to the non-equilibrium scaling of dissipation, i.e., $\varepsilon~\sim(\Rey_k/\Rey)^2k^{3/2}/L$ instead of the classical $\varepsilon \sim k^{3/2}/L$. The region where this non-equilibrium scaling emerges is marked by the appearance of a visible large-scale helical structure, indicating a link between the coherent structures and non-classical wake scalings.

For the stratified wakes, we find that the wake generator has a significant influence on the wake decay rates. There are substantial differences with the wakes of bluff bodies in similar conditions, particularly for the strongly stratified case of $\Fro = 2$. The $\Fro = 2$ wake shows $U_d \sim x^{-3/4}$ from $Nt > 15$. Although, the rate of decay is in accord with the quasi-two-dimensional (Q2D) power-law of Spedding \cite{Spedding1997}, it occurs quite early compared to previous experimental and numerical works where the Q2D regime appears around $Nt \approx 50$. Visualizations show that both mean and turbulence are strongly modulated by the steady lee waves and the unique separation pattern of the turbulent boundary layer. 

Meunier and Spedding \cite{Meunier2004} compared the evolution far into the  stratified wake, up to $x \approx 8000$, among  several body shapes. The body Reynolds number was $Re = 5000$. The defect velocity showed difference among the various bodies until  $Nt \approx 30$ but these differences reduced with increasing $Nt$ so that by $Nt \approx 50$ the laboratory data suggested a universal decay with the Q2D power-law exponent. In contrast, the present result for the $\Fro = 2$ wake shows an earlier onset of the Q2D regime at $Nt \approx 15$. The value of $Re = 10^5$ is  larger here and it is possible that features linked to the early onset, i.e.,  the instability that leads to horizontal meanders and also the enhanced TKE production, are inhibited by viscous damping at the lower $\Rey$ of the experiments. The present simulations do not extend into the very far wake reached in the experiments. Future  numerical or experimental work   at higher $\Rey$ that probes the very far wake would clearly be useful. To do so in simulations, it will be efficient to utilize the temporal model and initialize it with data from a hybrid simulation of the type conducted here \cite{Pasquetti2011}.

\section{Acknowledgments}
\label{SECacknowledgements}

We gratefully acknowledge the support of ONR grant N00014-20-1-2253. Computational resources were provided by the Department of Defense High Performance Computing Modernization Program. 

\bibliographystyle{unsrt}
\bibliography{references}

\begin{thebibliography}{10}

\bibitem{Wang1970}
K.~C. Wang.
\newblock {Three-dimensional boundary layer near the plane of symmetry of a
  spheroid at incidence}.
\newblock {\em J. Fluid Mech.}, 43:187--209, 1970.

\bibitem{Costis1989}
C.~E. Costis, D.~P. Telionis, and N.~T. Hoang.
\newblock {Laminar separating flow over a prolate spheroid}.
\newblock {\em J. Aircr.}, 26(9):810--816, 1989.

\bibitem{Wang1990}
K.~C. Wang, H.~C. Zhou, C.~H. Hu, and S.~Harrington.
\newblock {Three-dimensional separated flow structure over prolate spheroids}.
\newblock {\em Proc. R. Soc. London, Ser. A: Mathematical and Physical
  Sciences}, 429(1876):73--90, 1990.

\bibitem{Chesnakas1994}
C.~J. Chesnakas and R.~L. Simpson.
\newblock {Full three-dimensional measurements of the cross-flow separation
  region of a 6:1 prolate spheroid}.
\newblock {\em Exp. Fluids}, 17:68--74, 1994.

\bibitem{Fu1994}
T.~C. Fu, A.~Shekarriz, J.~Katz, and T.~T. Huang.
\newblock {The flow structure in the lee of an inclined 6:1 prolate spheroid}.
\newblock {\em J. Fluid Mech.}, 269:79--106, 1994.

\bibitem{Constantinescu2002}
G.~S. Constantinescu, H.~Pasinato, Y.~Q. Wang, J.~R. Forsythe, and K.~D.
  Squires.
\newblock {Numerical investigation of flow past a prolate spheroid}.
\newblock {\em J. Fluids Eng.}, 124(4):904--910, 2002.

\bibitem{Wikstrom2004}
N.~Wikstr{\"{o}}m, U.~Svennberg, N.~Alin, and C.~Fureby.
\newblock {Large eddy simulation of the flow around an inclined prolate
  spheroid}.
\newblock {\em J. Turbul.}, 5(29), 2004.

\bibitem{Chevray1968}
R.~Chevray.
\newblock {The turbulent wake of a body of revolution}.
\newblock {\em J. Basic Eng}, 90:275--284, 1968.

\bibitem{Jimenez2010}
J.~M. Jim{\'{e}}nez, M.~Hultmark, and A.~J. Smits.
\newblock {The intermediate wake of a body of revolution at high Reynolds
  numbers}.
\newblock {\em J. Fluid Mech.}, 659:516--539, 2010.

\bibitem{Posa2016}
A.~Posa and E.~Balaras.
\newblock {A numerical investigation of the wake of an axisymmetric body with
  appendages}.
\newblock {\em J. Fluid Mech.}, 792:470--498, 2016.

\bibitem{Kumar2018}
P.~Kumar and K.~Mahesh.
\newblock {Large-eddy simulation of flow over an axisymmetric body of
  revolution}.
\newblock {\em J. Fluid Mech.}, 853:537--563, 2018.

\bibitem{Ortiz2021}
J.~L. Ortiz-Tarin, S.~Nidhan, and S.~Sarkar.
\newblock {High-Reynolds number wake of a slender body}.
\newblock {\em J. Fluid Mech.}, 261:333--374, 2021.

\bibitem{Tennekes1972}
H.~Tennekes and J.~L. Lumley.
\newblock {\em {A First Course in Turbulence}}.
\newblock MIT Press, 1972.

\bibitem{George1989}
W.~K. George.
\newblock {The self-preservation of turbulent flows and its relation to initial
  conditions and coherent structures}.
\newblock In {\em Advances in Turbulence}. Springer, 1989.

\bibitem{Pope2000}
S.~B. Pope.
\newblock {\em Turbulent Flows}.
\newblock Cambridge University Press, 2000.

\bibitem{Spedding2014}
G.~R. Spedding.
\newblock {Wake Signature Detection}.
\newblock {\em Annu. Rev. Fluid Mech.}, 46(1):273--302, 2014.

\bibitem{Gourlay2001}
M.~J. Gourlay, S.~C. Arendt, D.~C. Fritts, and J.~Werne.
\newblock {Numerical modeling of initially turbulent wakes with net momentum}.
\newblock {\em Phys. Fluids}, 13(12):3783--3802, 2001.

\bibitem{Dommermuth2002}
D.~G. Dommermuth, J.~W. Rottman, G.~E. Innis, and E.~A. Novikov.
\newblock {Numerical simulation of the wake of a towed sphere in a weakly
  stratified fluid}.
\newblock {\em J. Fluid Mech.}, 473(473):83--101, 2002.

\bibitem{Brucker2010}
K.~A. Brucker and S.~Sarkar.
\newblock {A comparative study of self-propelled and towed wakes in a
  stratified fluid}.
\newblock {\em J. Fluid Mech.}, 652:373--404, 2010.

\bibitem{Ortiz-tarin2019}
J.~L. Ortiz-Tarin, K.~C. Chongsiripinyo, and S.~Sarkar.
\newblock {Stratified flow past a prolate spheroid}.
\newblock {\em Phys. Rev. Fluids}, 094803:1--28, 2019.

\bibitem{nidhan2019dynamic}
S.~Nidhan, J.~L. Ortiz-Tarin, K.~Chongsiripinyo, S.~Sarkar, and P.~J. Schmid.
\newblock Dynamic mode decomposition of stratified wakes.
\newblock In {\em AIAA Aviation 2019 Forum}, page 3330, 2019.

\bibitem{Chongsiripinyo2020}
K.~Chongsiripinyo and S.~Sarkar.
\newblock {Decay of turbulent wakes behind a disk in homogeneous and stratified
  fluids}.
\newblock {\em J. Fluid Mech.}, 885, 2020.

\bibitem{Nidhan2020}
S.~Nidhan, K.~Chongsiripinyo, O.~T. Schmidt, and S.~Sarkar.
\newblock {Spectral POD analysis of the turbulent wake of a disk at Re=50000}.
\newblock {\em Phys. Rev. Fluids}, 2020.

\bibitem{nidhan2022analysis}
S.~Nidhan, O.~T. Schmidt, and S.~Sarkar.
\newblock Analysis of coherence in turbulent stratified wakes using spectral
  proper orthogonal decomposition.
\newblock {\em J. Fluid Mech.}, 934, 2022.

\bibitem{VanDine2018}
A.~VanDine, K.~Chongsiripinyo, and S.~Sarkar.
\newblock Hybrid spatially-evolving {DNS} model of flow past a sphere.
\newblock {\em Comput. Fluids}, 171:41--52, 2018.

\bibitem{Pasquetti2011}
R.~Pasquetti.
\newblock {Temporal/spatial simulation of the stratified far wake of a sphere}.
\newblock {\em Comput. Fluids}, 40(1):179--187, 2011.

\bibitem{Nicoud1999}
F.~Nicoud and F.~Ducros.
\newblock Subgrid-scale stress modelling based on the square of the velocity
  gradient tensor.
\newblock {\em Flow Turbul. Combust.}, 62:183--200, 1999.

\bibitem{Balaras2004}
E.~Balaras.
\newblock Modeling complex boundaries using an external force field on fixed
  {C}artesian grids in large-eddy simulations.
\newblock {\em Comput. Fluids}, 33:375--404, 03 2004.

\bibitem{Nedic2013}
J.~Nedi{\'{c}}, J.~C. Vassilicos, and B.~Ganapathisubramani.
\newblock {Axisymmetric turbulent wakes with new nonequilibrium similarity
  scalings}.
\newblock {\em Phys. Rev. Lett.}, 111(14):1--5, 2013.

\bibitem{Dairay2015}
T.~Dairay, M.~Obligado, and J.~C. Vassilicos.
\newblock {Non-equilibrium scaling laws in axisymmetric turbulent wakes}.
\newblock {\em J. Fluid Mech.}, 781:166--195, 2015.

\bibitem{Cafiero2019}
G.~Cafiero and J.~C. Vassilicos.
\newblock {Non-equilibrium turbulence scalings and self-similarity in turbulent
  planar jets}.
\newblock {\em Proc. Royal Soc. A}, 2019.

\bibitem{Goto2016}
S.~Goto and J.~C. Vassilicos.
\newblock {Unsteady turbulence cascades}.
\newblock {\em Phys. Rev. E}, 94(5):1--3, 2016.

\bibitem{Berger1990}
E.~Berger, D.~Scholz, and M.~Schumm.
\newblock {Coherent vortex structures in thewake of a sphere and a circular
  disk at rest and under forced vibrations}.
\newblock {\em J. Fluids Struct.}, 4(3):231--257, 1990.

\bibitem{Johansson2006}
P.~B.~V. Johansson and W.~K. George.
\newblock {The far downstream evolution of the high-Reynolds-number
  axisymmetric wake behind a disk. Part 1. Single-point statistics}.
\newblock {\em J. Fluid Mech.}, 555:363--385, 2006.

\bibitem{Hanazaki1988}
H.~Hanazaki.
\newblock {A numerical study of three-dimensional stratified flow past a
  sphere}.
\newblock {\em J. Fluid Mech.}, 192(1988):393--419, 1988.

\bibitem{Chomaz1992}
J.~M. Chomaz, P.~Bonneton, A.~Butet, M.~Perrier, and E.~J. Hopfinger.
\newblock {Froude number dependence of the flow separation line on a sphere
  towed in a stratified fluid}.
\newblock {\em Phys. Fluids}, 4(2):254--258, 1992.

\bibitem{Spedding1997}
G.~R. Spedding.
\newblock {The evolution of initially turbulent bluff-body wakes at high
  internal Froude number}.
\newblock {\em J. Fluid Mech.}, 337:283--301, 1997.

\bibitem{Diamessis2011}
P.~J. Diamessis, G.~R. Spedding, and J.~A. Domaradzki.
\newblock {Similarity scaling and vorticity structure in high-Reynolds-number
  stably stratified turbulent wakes}.
\newblock {\em J. Fluid Mech.}, 671:52--95, 2011.

\bibitem{Redford2015}
J.~A. Redford, T.~S. Lund, and G.~N. Coleman.
\newblock {A numerical study of a weakly stratified turbulent wake}.
\newblock {\em J. Fluid Mech.}, 776:568--609, 2015.

\bibitem{Meunier2004}
P.~Meunier and G.~R. Spedding.
\newblock {A loss of memory in stratified momentum wakes}.
\newblock {\em Phys. Fluids}, 16(2):298--305, 2004.

\end{thebibliography}

\end{multicols*}
\end{document}